Lan Chu Khanh

Research and Consultancy Department, Banking Academy of Vietnam

Hanoi 10000, Vietnam

lanck@hvnh.edu.vn

# The effects of institutional quality on formal and informal borrowing across high-, middle-, and low-income countries

**Abstract**

This paper examines the effects of institutional quality on financing choice of individual using a large dataset of 137,160 people from 131 countries. We classify borrowing activities into three categories, including formal, constructive informal, and underground borrowing. Although the result shows that better institutions aids the uses of formal borrowing, the impact of institutions on constructive informal and underground borrowing among three country sub-groups differs. Higher institutional quality improves constructive informal borrowing in middle-income countries but reduces the use of underground borrowing in high- and low-income countries.

*Keywords:* financial inclusion, formal borrowing, informal borrowing, institutional quality

*JEL classification:* G21, O17, O43

## 1. Introduction

According to World Bank's Global Findex Database Report, globally, around 1.7 billion adults remain unbanked. Moreover, while nearly 60% of borrowers in high-income countries reported borrowing from a financial institution or using a credit card, the ratio in developing countries is about one third of such number. This urges for the research on which and how individual and country factors determine inclusive finance in general, and borrowing practices in particular. The objective of this paper is to identify how institutional quality affects the practice of household's formal and informal borrowing across high-, middle-, and low-income countries. We refer a loan as a formal borrowing if the lender is a bank or another type of financial institution and as an informal borrowing otherwise. We follow Allen et al. (2018) to classify informal borrowing into two categories, namely constructive informal borrowing and underground borrowing. The former includes trade credit from a store, lending from relatives or friends; thus, to some extent, it is deemed to benefit the borrower, as there exists the exchange of information between two parties. In contrast, underground financing, i.e. credit from loan sharks, unregistered pawnshop and so on, has characteristics of little information exchange, high interest rate, the high probability of violence in case of delinquency. We apply these definitions to classify respondents in the World Bank's 2017 Global Financial Inclusion database into three types of who borrows from formal, constructive informal, and underground borrowing sources. Next, to investigate the impact of institutional quality on financial inclusion, we collect the institutional quality indicators from World Governance Indicator database.

This paper fills the literature gap of determinants, especially institutional quality, of financial inclusion. Overall, we find that being a male, more educated, richer, and older to a certain age has a higher probability of borrowing formally. However, constructive informal financing, like borrowing from family, friends, or a store, is not significantly affected by education level in all three country sub-groups while underground financing activities is better controlled when people are more educated in high- and middle-income countries only. The positive role of higher income level in reducing the likelihood of borrowing informally occurs mostly in high-income countries. Our first key finding on the effect of institutional quality is that better institutional quality favors the prevalence of formal borrowing in all three country sub-groups. Second, higher institutional quality supports the use of constructive informal borrowing in middle-income countries. For underground financing, better institutional quality significantly decreases the likelihood that a person in high- and low-income countries takes out an underground loan.

The rest of the paper is organized as follows. Section 2 reviews related literature. Section 3 details the data and empirical approach used. Section 4 presents results. Section 5 concludes.

## 2. Literature review

There are two popular strands of literature on the relationship between institutional quality and financial development as well as the determinants of financial inclusion. In this section, we briefly review the development of these two groups.

In recent years, institutional quality has gained popularity in explaining the finance development and economic growth relationship. Using panel data of 72 countries from 1978 to 2000, Demetriades and Law (2006) conclude that financial development has larger effect on long-run economic development when the financial system is embedded within a sound institutional framework. They use the institutions dataset from Internal Country Risk Guide, published by Political Risk Services, which cover five indicators, namely, corruption, law and order, bureaucratic quality, government stability, and democracy and accountability. Law et al. (2013) find that the marginal effect of financial development on growth depends on institutional quality. The institutional quality is measured using two datasets, the Internal Country Risk Guide and World Governance Indicators. The estimation result for 87 countries shows that the finance curse occurs under weak institution quality conditions. Countries with better institutions receive more benefits from financial development than low-institutional quality countries do. Slesman et al. (2019) argue that there exists a critical threshold of political institution, over which financial development has a positive effect on economic growth. Based on a dataset of 77 developing and emerging countries, they find robust evidences that good-quality political institution is an important conductor for financial development to foster economic growth. Institutions also influence poverty and inequality. Chong and Calderon (2000) concluded that high institutional quality reduces inequality in rich countries but increases inequality in poor countries. Chong and Gradstein (2007) use Internal Country Risk Guide dataset to measure the impact of institutions on inequality in more than 100 countries. They find that institution quality is an important tool for reducing inequality. Majeed (2017) concludes that both inclusive financial development and institutions are essential for eradicating poverty for Islamic countries.

The second line of literature studies the determinants of financial inclusion as well as its impacts on household living and firm's operation. Allen et al. (2016) study the factors underpinning the account ownership and usage. They take advantage of 2014 Global Financial Inclusion database and a variety of country level characteristics to find how different individual and country characteristics affect the the probability of owing an account, using it to save, and using it frequently. The results show that greater inclusive finance associates with better environment to access financial services. Zins and Weill (2016) find that being a man, richer, more educated, and older to a certain age increases the likelihood of using formal financial services in African countries. Fungáčová and Weill (2015) find the similar results for Chinese people. Recent research also investigate the factors determining informal financing, the relationship between formal financing and informal financing, as well as the impacts of informal financing on both individuals and firms. Khoi et al. (2013) examine the factors influencing household's access to formal and informal credit in the Vietnamese rural market. Using data surveyed from nearly 1,000 households from 15 villages of 13 communes in the Mekong River Delta, they find the interaction between the two credit sectors in which informal credit can alter the likelihood of the households' access to formal microcredit program. They also indicate several determinants of informal borrowing, such as land holding, loan purpose, interest rate, loan duration, and road access. Using household panel dataset in rural Bangladesh, Islam et al. (2014) indicate that access to microfinance reduces the probability of taking out an informal loan but not the loan amount. For firms, Dygryse et al. (2016) conclude that different types of external finance have different beneficial impacts on the borrowing firms. They identify a complementary effect between formal and informal finance for the sales growth of small firms but not for large firms. Allen et al. (2108) divide informal financing into two categories with distinct characteristics based on two criteria, information technology for monitoring, risk control, pricing and the violence mechanism in case of delinquency. Using data from China's small and medium enterprises, they find that constructive informal financing supports firm growth while underground financing do not.

Recently, several researchers connect two above-mentioned literature lines to examine the effect of institutional quality on financial inclusion. Madestam (2014) develops a model to investigate whether and how legal institutions affect the prevalence of informal finance. It is confirmed that weaker legal institutions increase the popularity of informal credit if borrowers obtain money from both financial sectors while the opposite is true if informal lenders supply all funds. Allen et al. (2016) investigate the impacts of legal rights index and political risk rating on the ownership and usage of account. They find that both indicators positively affect the financial inclusion indicator after controlling for level of economic development. However, the literature is still lack of research examining the effect of institutional quality on informal and formal borrowing. Our paper deals with this gap by investigating impact of institutional quality, proxied by government effectiveness, regulatory quality, and rule of law, on how individual borrows.

## 3. Data and model specification

### 3.1. Data

We utilize the World Bank's Governance Indicators Database and 2014 Global Financial Inclusion Database to measure the effects of institutional quality on individual borrowing activities. The Global Financial Inclusion Database is the most comprehensive database that reflects how adults around the world save, borrow, make payments, and manage risks. The database is collected through the 2014 calendar year, covering about 150,000 adults in more than 140 economies. In this paper, we focus only on the financing side of financial inclusion, which includes both formal and informal borrowing activities. Formal borrowing is defined as an individual borrows any money from a bank, or another type of formal financial institution. We follow Allen et al. (2018) to classify informal borrowing into two categories, namely constructive informal borrowing and underground borrowing. The formal includes activities of borrowing from a store using installment credit, buying on credit, and borrowing from family, relatives, or friends. This informal sources use personal, community, or business relationships to reduce asymmetric information and risk through economic collateral. The price of credit reflects both the risk and the closeness of the relationship between borrowers and lenders (Allen et al. 2018). In contrast, the latter category refers to respondents who borrow from another private lender, i.e., loan shark, payday lender, or pawnshop. This lending practice associates with low information level exchange, high lending rate and fee, and sometimes, violence in the case of delinquency. Thus, we have four variables, including formal borrowing, informal borrowing, informal constructive borrowing, and underground borrowing. The four indicators are converted to dummy variables which equal to one if the respondent responded yes to the surveyed question and zero otherwise.

Individual characteristics that affect borrowing activities are gender, age, education, and income. Gender is a dummy variable equal to one if the individual is a man and zero otherwise. To consider the possible non-linear relationship between age and financial inclusion mentioned in previous research, we use both individual's age and its squared. Education level is measured by two dummy variables, which equal to one if the individual has secondary education and completed tertiary education or more. We use four dummy variables to measure income (second 20%, third 20%, fourth 20%, and richest 20%) which equal to one if the individual's income belongs to the corresponding quintile. The omitted dummy variable is for the poorest 20%. All these explanatory variables are provided in the survey dataset published by the World Bank.

We follow the World Bank (2002) to define institutional, or governance quality, as "the traditions and institutions by which authority in a country is exercised". The six dimensions of institutional quality belongs to three areas, includes "the process by which governments are selected, monitored, and replaced", "the capacity of the government to effectively formulate and implement sound policies", and "the respect of citizens and the stage for the institutions that govern economic and social interactions among them". However, in this paper, we employ only three out of six indicators, namely, government effectiveness and regulatory quality of the second area as well as rule of law of the third area to represent the institutional quality[1]. We select these indicators because they are more relevant in determining financial inclusion than other remaining indicators. The first indicator, government effectiveness, measures the quality of public service, civil service, as well as policy formulation and implementation. These factors are very important in ensuring that all citizens are able to receive the benefit of public services, which is a pre-determinant for accessing and using financial services. The second indicator, regulatory quality, captures the ability of government to conduct policies and regulations that promotes private sector development. The development of private sector is one of determining factors that push the inclusive finance in both supply and demand sides. The third indicator, rule of law, quantifies the extent to which agents have confidence in and abide by the rules of society. This indicator is deemed to strongly affect financial inclusion as it reflects the quality of contract enforcement, property rights, the police, and the courts, as well as the likelihood of crime and violence. Although we average the three indicators to obtain one institutional quality indicator, we conduct regressions for all three indicators to see the individual effect of each indicator on financing choices. We take institutional quality data in the year 2013, back to 1 year in compared to the year of financial inclusion data.

---

[1] For a detailed discussion of institutional quality, see Kaufmann et al. (2010).

Beside the individual and country characteristics mentioned above, we also include four variables that reflect the socio-economic condition of an economy. First, the variable GDP per capita is used to capture the effect of different level of economic development. Second, we use the access to electricity in rural area to measure the infrastructure level. Third, we use population density to measure the geographic potential of the country and the size of the economy. Fourth, the availability of banking services in a country, measured by the number of commercial bank branches per 100,000 adults, is employed.

### 3.2. Model specification

We perform probit estimations to explain the determinants of borrowing practice using the following equation:

$$y_{i,j} = \alpha + \beta * x'_{i,j} + \gamma * z'_{i,j} + \varepsilon_{i,j} \tag{1}$$

where, countries and individuals are indexed by i and j; the dependent variable $y_{i,j}$ represents four measures of borrowing practices, including formal borrowing, informal borrowing, informal constructive borrowing, and underground borrowing; $x'_{i,j}$ is a vector of individual characteristics, $z'_{i,j}$ is a vector of country characteristics, including institutional quality; $\varepsilon_{i,j}$ is a normally distributed error term with zero mean and variance equals to 1. As the primary goal of this paper is to compare the determinants of formal and informal financing for different groups of countries, we run model (1) with full sample and three country sub-groups according to World Bank's classification. The standard errors are robust and clustered at the country level.

### 4. Results

### 4.1. Descriptive results

Table 1 presents the main descriptive statistics for four borrowing variables for the global level and three groups of countries. We find that the practice of formal borrowing increases with the economic development level. Nearly one of five surveyed respondents answer that they borrow money from a bank or another financial institution while the ratio in middle-income and low-income countries is one-eight and one-seventeen respectively. In contrast, the practice of informal borrowing is more popular in lower income countries. One of four surveyed respondents report that they borrowing informally in high-income countries. Moreover, only 2% of respondents in this country group answer that they do take out an underground borrowing in the past year. This ratio increases 2.5 times to 5% in middle- and low-income countries.

**Table 1: Indicators for formal and informal borrowing: full and split samples**

|  | Full sample | | | High-income | | | Middle-income | | | Low-income | | |
| --- | --- | --- | --- | --- | --- | --- | --- | --- | --- | --- | --- | --- |
|  | Obs. | Mean | Std. dev | Obs. | Mean | Std. dev | Obs. | Mean | Std. dev | Obs. | Mean | Std. dev |
| Formal borrowing | 137,160 | 0.130 | 0.336 | 45,170 | 0.174 | 0.379 | 71,400 | 0.123 | 0.328 | 20,590 | 0.059 | 0.236 |
| Informal borrowing | 137,160 | 0.306 | 0.461 | 45,170 | 0.240 | 0.427 | 71,400 | 0.318 | 0.466 | 20,590 | 0.412 | 0.492 |
| Constructive borrowing | 137,160 | 0.294 | 0.456 | 45,170 | 0.235 | 0.424 | 71,400 | 0.303 | 0.459 | 20,590 | 0.398 | 0.489 |
| Underground borrowing | 137,160 | 0.040 | 0.197 | 45,170 | 0.020 | 0.141 | 71,400 | 0.050 | 0.218 | 20,590 | 0.052 | 0.221 |

Table 2 presents the main descriptive statistics for four measures of institutional quality for the global level and three groups of countries. The estimate of institutional quality is a standard normal random variable, with zero mean, unit standard deviation, and ranging approximately from -2.5 to 2.5. We find that there are large differences in institutional quality for three country sub-groups. As expected, countries in high-income have highest scores of overall institutional quality, and all three components, followed by middle-income and low-income countries.

**Table 2: Indicators for institutional quality: full and split samples**

|  | Full sample | | | High-income | | | Middle-income | | | Low-income | | |
| --- | --- | --- | --- | --- | --- | --- | --- | --- | --- | --- | --- | --- |
|  | Obs. | Mean | Std. dev | Obs. | Mean | Std. dev | Obs. | Mean | Std. dev | Obs. | Mean | Std. dev |
| Overall institutional quality | 131 | 0.144 | 0.915 | 45 | 1.032 | 0.594 | 65 | -0.419 | 0.501 | 21 | -0.825 | 0.425 |
| Government effectiveness | 131 | 0.090 | 0.876 | 45 | 1.184 | 0.581 | 65 | -0.340 | 0.527 | 21 | -0.927 | 0.421 |
| Regulatory quality | 131 | 0.152 | 0.953 | 45 | 1.150 | 0.604 | 65 | -0.253 | 0.637 | 21 | -0.735 | 0.453 |
| Rule of law | 131 | 0.006 | 1.005 | 45 | 1.140 | 0.648 | 65 | -0.499 | 0.532 | 21 | -0.859 | 0.478 |

### 4.2. Main results

Table 3 presents estimation result of determinants of formal and informal borrowing. In the left panel of Table 3, we find that a male, more educated, richer, and older to a certain age has a higher probability of taking out a loan from financial institution. The marginal probability increases with education and income level in full sample and all three country sub-groups. Better infrastructure development, lower population density, and higher institutional quality associate with higher level of financial inclusion in full sample. Population density has different impact on formal borrowing in high-income and low-income countries. In the former, lower population densitiy significantly increases the practice of formal borrowing. In contrast, increasing population density encourages formal borrowing activities in low-income countries. Improving infrastructure is a key to develop inclusive finance in middle-income countries.

For informal borrowing, we find that gender and age, but not education, are important factors in determining the likelihood that an individual can obtain an informal loan. While in high-income countries, an individual belongs to all four income-quintiles has the lower likelihood of taking out an informal loan, it is only the case of individual who belongs to the the highest income-quintile in middle-income countries. We find that better access to electricity, prevalent commercial branches, and high level of institutions significantly reduce the probability that an individual takes out an informal loan in middle-income countries.

**Table 3: Effects of institutional quality on formal and informal borrowing**

|  | Formal borrowing | | | | Informal borrowing | | | |
|---|---|---|---|---|---|---|---|---|
|  | Full sample | High-income | Middle-income | Low-income | Full sample | High-income | Middle-income | Low-income |
| Male | 0.108*** | 0.130*** | 0.091*** | 0.109* | 0.068*** | 0.058** | 0.068*** | 0.064*** |
|  | (0.015) | (0.022) | (0.022) | (0.061) | (0.012) | (0.024) | (0.016) | (0.021) |
| Age | 0.082*** | 0.078*** | 0.087*** | 0.066*** | 0.038*** | 0.014** | 0.046*** | 0.043*** |
|  | (0.004) | (0.007) | (0.005) | (0.010) | (0.003) | (0.006) | (0.004) | (0.005) |
| Age_squared | -0.001*** | -0.001*** | -0.001*** | -0.001*** | -0.001*** | -0.000*** | -0.001*** | -0.001*** |
|  | (0.000) | (0.000) | (0.000) | (0.000) | (0.000) | (0.000) | (0.000) | (0.000) |
| Secondary education | 0.191*** | 0.148*** | 0.199*** | 0.210*** | 0.001 | 0.019 | 0.018 | 0.027 |
|  | (0.036) | (0.046) | (0.052) | (0.074) | (0.032) | (0.042) | (0.043) | (0.086) |
| Tertiary education | 0.342*** | 0.261*** | 0.407*** | 0.380*** | -0.033 | -0.037 | 0.056 | -0.247 |
|  | (0.042) | (0.059) | (0.055) | (0.092) | (0.040) | (0.055) | (0.050) | (0.190) |
| Income-second 20% | 0.093*** | 0.090*** | 0.113*** | 0.026 | -0.006 | -0.079** | 0.023 | 0.080* |
|  | (0.022) | (0.027) | (0.034) | (0.093) | (0.021) | (0.033) | (0.030) | (0.048) |
| Income-third 20% | 0.158*** | 0.099*** | 0.200*** | 0.228*** | -0.029 | -0.123*** | 0.007 | 0.073 |
|  | (0.022) | (0.032) | (0.030) | (0.066) | (0.024) | (0.036) | (0.034) | (0.068) |
| Income-fourth 20% | 0.180*** | 0.163*** | 0.211*** | 0.132 | -0.056** | -0.144*** | -0.024 | 0.032 |
|  | (0.023) | (0.034) | (0.032) | (0.083) | (0.025) | (0.035) | (0.036) | (0.070) |
| Income-richest 20% | 0.213*** | 0.118*** | 0.258*** | 0.365*** | -0.126*** | -0.225*** | -0.104*** | -0.008 |
|  | (0.029) | (0.036) | (0.044) | (0.069) | (0.024) | (0.037) | (0.035) | (0.054) |
| GDP per capita | -0.076* | 0.001 | -0.085 | 0.040 | -0.082* | 0.042 | -0.062 | -0.345 |
|  | (0.042) | (0.074) | (0.082) | (0.221) | (0.050) | (0.099) | (0.077) | (0.231) |
| Access to electricity | 0.111*** | 0.088 | 0.090* | 0.035 | -0.049 | 0.062 | -0.131** | -0.061 |
|  | (0.033) | (0.224) | (0.054) | (0.040) | (0.039) | (0.415) | (0.056) | (0.073) |
| Population density | -0.044* | -0.069** | -0.038 | 0.192*** | -0.001 | -0.047 | 0.017 | -0.022 |
|  | (0.026) | (0.031) | (0.042) | (0.060) | (0.025) | (0.037) | (0.038) | (0.097) |
| Commercial branches | 0.022 | -0.033 | 0.031 | -0.014 | -0.064 | 0.012 | -0.122** | 0.139 |
|  | (0.048) | (0.056) | (0.060) | (0.118) | (0.041) | (0.080) | (0.052) | (0.214) |
| Institutional quality | 0.229*** | 0.191*** | 0.239*** | 0.233* | 0.084 | -0.159 | 0.224* | 0.190 |
|  | (0.049) | (0.060) | (0.082) | (0.138) | (0.065) | (0.145) | (0.125) | (0.225) |
| Constant | -2.808*** | -2.979*** | -2.825*** | -4.192*** | 0.043 | -0.885 | 0.121 | 1.452 |
|  | (0.368) | (0.834) | (0.742) | (1.402) | (0.426) | (1.874) | (0.701) | (1.578) |
| Pseudo R-squared | 0.075 | 0.057 | 0.064 | 0.076 | 0.035 | 0.046 | 0.032 | 0.021 |
| Log likelihood | -46,346 | -18,126 | -24,637 | -3,363 | -80,633 | -22,845 | -45,517 | -11,628 |
| Observations | 133,836 | 41,936 | 74,327 | 17,573 | 133,836 | 41,936 | 74,327 | 17,573 |

\*\*\*, \*\*, \* indicate significance at the 1%, 5%, and 10% levels respectively. Clustered errors are in parentheses.

Table 4 shows that the effects of three components on each country sub-group are different. In high-income and middle-income country sub-groups, government effectiveness, regulatory quality, and rule of law have statistically significant and positive influence on formal borrowing. In low-income country, only rule of law significantly determines formal borrowing. While government effectiveness has the highest impact on formal borrowing in high-income countries, regulatory quality is the most important factor in middle-income countries. The rule of law has the highest impact on formal borrowing in low-income countries and among three country sub-groups.

**Table 4: Effects of institutional quality components on formal and informal borrowing**

|  | Formal borrowing | | | | Informal borrowing | | | |
|---|---|---|---|---|---|---|---|---|
|  | Full sample | High-income | Middle-income | Low-income | Full sample | High-income | Middle-income | Low-income |
| Government effectiveness | 0.206*** | 0.195*** | 0.178** | 0.194 | 0.075 | -0.146 | 0.160 | 0.244 |
|  | (0.054) | (0.061) | (0.086) | (0.140) | (0.066) | (0.119) | (0.117) | (0.220) |
| Regulatory quality | 0.239*** | 0.184*** | 0.265*** | 0.187 | 0.098 | -0.102 | 0.249** | 0.053 |
|  | (0.048) | (0.053) | (0.079) | (0.130) | (0.063) | (0.139) | (0.105) | (0.219) |
| Rule of law | 0.161*** | 0.143* | 0.133* | 0.253* | 0.049 | -0.201 | 0.134 | 0.239 |
|  | (0.047) | (0.074) | (0.077) | (0.131) | (0.056) | (0.145) | (0.112) | (0.189) |

\*\*\*, \*\*, \* indicate significance at the 1%, 5%, and 10% levels respectively. Clustered errors are in parentheses.

Table 5 examines the determinants of two types of informal borrowing. In the left panel of Table 5, we find that the results are quite similar to the results of informal borrowing. The right panel of Table 5 shows some notable results. First, higher education level significantly reduces the probability that an individual takes out an underground loan in high-income and middle-income countries. Greater income negatively associates with higher underground borrowing in full sample and high-income countries only. Better infrastructure in high-income countries, more commercial bank branches in high- and midlle-income countries, and higher population density in low-income countries associate with lower practice of underground borrowing. It is also noted that the positive effect of institutional quality on informal borrowing in middle-income countries found in Table 4 comes from the constructive informal borrowing rather than underground borrowing.

**Table 5: Effects of institutional quality on constructive informal and underground borrowing**

|  | Constructive informal borrowing | | | | Underground borrowing | | | |
| --- | --- | --- | --- | --- | --- | --- | --- | --- |
|  | Full sample | High-income | Middle-income | Low-income | Full sample | High-income | Middle-income | Low-income |
| Male | 0.061*** | 0.055** | 0.063*** | 0.047*** | 0.105*** | 0.067 | 0.094*** | 0.163*** |
|  | (0.012) | (0.024) | (0.016) | (0.018) | (0.019) | (0.054) | (0.023) | (0.034) |
| Age | 0.037*** | 0.014** | 0.044*** | 0.043*** | 0.043*** | 0.028*** | 0.052*** | 0.023*** |
|  | (0.003) | (0.006) | (0.003) | (0.005) | (0.004) | (0.009) | (0.004) | (0.008) |
| Age_squared | -0.001*** | -0.000*** | -0.001*** | -0.001*** | -0.001*** | -0.000*** | -0.001*** | -0.000*** |
|  | (0.000) | (0.000) | (0.000) | (0.000) | (0.000) | (0.000) | (0.000) | (0.000) |
| Secondary education | 0.008 | 0.024 | 0.029 | 0.022 | -0.091* | -0.111 | -0.112* | -0.051 |
|  | (0.031) | (0.044) | (0.040) | (0.084) | (0.053) | (0.073) | (0.064) | (0.086) |
| Tertiary education | -0.025 | -0.028 | 0.069 | -0.251 | -0.132** | -0.169* | -0.154* | -0.014 |
|  | (0.039) | (0.057) | (0.047) | (0.181) | (0.066) | (0.088) | (0.081) | (0.133) |
| Income-second 20% | -0.005 | -0.074** | 0.020 | 0.089* | -0.044 | -0.161*** | -0.010 | -0.022 |
|  | (0.021) | (0.033) | (0.029) | (0.048) | (0.028) | (0.059) | (0.036) | (0.053) |
| Income-third 20% | -0.032 | -0.125*** | -0.000 | 0.083 | -0.002 | -0.109 | 0.033 | -0.004 |
|  | (0.024) | (0.036) | (0.033) | (0.068) | (0.033) | (0.072) | (0.044) | (0.068) |
| Income-fourth 20% | -0.053** | -0.138*** | -0.026 | 0.049 | -0.069* | -0.228*** | -0.030 | -0.023 |
|  | (0.025) | (0.035) | (0.036) | (0.066) | (0.036) | (0.057) | (0.049) | (0.082) |
| Income-richest 20% | -0.123*** | -0.222*** | -0.104*** | 0.007 | -0.108*** | -0.285*** | -0.052 | -0.109 |
|  | (0.024) | (0.037) | (0.035) | (0.051) | (0.037) | (0.072) | (0.047) | (0.067) |
| GDP per capita | -0.080 | 0.021 | -0.053 | -0.351 | -0.017 | 0.509*** | -0.070 | -0.239 |
|  | (0.049) | (0.098) | (0.076) | (0.231) | (0.075) | (0.127) | (0.113) | (0.244) |
| Access to electricity | -0.056 | 0.530 | -0.137** | -0.068 | 0.057 | -3.105*** | 0.004 | 0.020 |
|  | (0.039) | (0.407) | (0.053) | (0.076) | (0.044) | (0.515) | (0.054) | (0.052) |
| Population density | -0.004 | -0.046 | 0.013 | -0.038 | 0.053 | -0.033 | 0.043 | 0.227*** |
|  | (0.025) | (0.037) | (0.037) | (0.101) | (0.041) | (0.054) | (0.050) | (0.055) |
| Commercial branches | -0.061 | 0.017 | -0.119** | 0.163 | -0.087 | -0.292** | -0.125* | 0.025 |
|  | (0.041) | (0.080) | (0.051) | (0.222) | (0.059) | (0.134) | (0.066) | (0.110) |
| Institutional quality | 0.097 | -0.130 | 0.229* | 0.229 | -0.195** | -0.667*** | 0.030 | -0.317** |
|  | (0.065) | (0.140) | (0.125) | (0.229) | (0.089) | (0.148) | (0.158) | (0.146) |
| Constant | 0.045 | -2.885 | 0.078 | 1.527 | -2.531*** | 8.451*** | -1.821* | -1.804 |
|  | (0.423) | (1.835) | (0.692) | (1.610) | (0.589) | (1.788) | (0.981) | (1.415) |
| Pseudo R-squared | 0.033 | 0.044 | 0.031 | 0.023 | 0.049 | 0.175 | 0.033 | 0.037 |
| Log likelihood | -79,560 | -22,650 | -44,780 | -11,508 | -21,612 | -3,362 | -14,237 | -3,422 |
| Observations | 133,836 | 41,936 | 74,327 | 17,573 | 133,836 | 41,936 | 74,327 | 17,573 |

***, **, * indicate significance at the 1%, 5%, and 10% levels respectively. Clustered errors are in parentheses.

Higher institutional quality significantly decreases the likelihood that an individual borrows from underground sources in high-income and low-income countries. Table 6 presents the effect of each component of institutional quality on constructive informal and underground borrowing. We find that in high-income countries, all three components prove effective in controlling the underground borrowing practice with the highest impact belongs to government effectiveness. In low-income countries, government effectiveness and regulator quality significantly decreases the prevalent of underground borrowing. In middle-income countries, we do not find any component that significantly associates with underground borrowing.

**Table 6: Effects of institutional quality components on constructive informal and underground borrowing**

|  | Constructive informal borrowing | | | | Underground borrowing | | | |
|---|---|---|---|---|---|---|---|---|
|  | Full sample | High-income | Middle-income | Low-income | Full sample | High-income | Middle-income | Low-income |
| Government effectiveness | 0.092 | -0.113 | 0.172 | 0.285 | -0.235** | -0.702*** | -0.045 | -0.333* |
|  | (0.066) | (0.117) | (0.116) | (0.218) | (0.095) | (0.110) | (0.154) | (0.195) |
| Regulatory quality | 0.106* | -0.080 | 0.246** | 0.085 | -0.126 | -0.511*** | 0.098 | -0.377*** |
|  | (0.063) | (0.133) | (0.105) | (0.227) | (0.079) | (0.158) | (0.130) | (0.135) |
| Rule of law | 0.061 | -0.176 | 0.137 | 0.274 | -0.162* | -0.668*** | 0.005 | -0.169 |
|  | (0.056) | (0.141) | (0.111) | (0.194) | (0.086) | (0.184) | (0.140) | (0.119) |

\*\*\*, \*\*, \* indicate significance at the 1%, 5%, and 10% levels respectively. Clustered errors are in parentheses.

## 5. Conclusion

In this paper, we examine the impacts of institutional quality on formal and informal borrowing based on sample 137,160 people from 131 countries. We find that high institutional quality significantly increases the probability that an individual takes out a formal loan. However, impacts of institutions on informal constructive and underground borrowing among country sub-groups are different. Higher institutional quality significantly associates with the constructive informal borrowing in middle-income countries while it decreases the likelihood that an individual borrows from underground sources in high and low-income countries.

For institutions components, we find that for high-income countries, government effectiveness, regulatory quality, and rule of law have statistically significant and positive impact on both formal borrowing and underground borrowing. In middle-income countries, all three components encourage formal borrowing practices but none of them proves effective in controlling underground borrowing. In low-income countries, government effectiveness and regulator quality significantly decreases the prevalent of underground borrowing.

For individual characteristics, we find that being a male, more educated, richer, and older to a certain age has a higher probability of borrowing formally. However, constructive informal financing is not significantly affect by education level in all three country sub-groups while underground financing activities is better controlled when people are more educated in only high- and middle-income countries. The positive role of higher income level in reducing the likelihood of borrowing informally occurs mostly in high-income countries.

The role of institutional quality found in this paper may have implications for policymakers in conducting policies for inclusive finance. Encouraging formal and constructive informal borrowing through institutional quality in three country sub-groups requires different measures. While policymakers in high- and middle-income countries can target policies that improve government effectiveness, regulatory quality, and rule of law, those in low-income countries should focus on the latter. In contrast, to control underground borrowing, those in low-income countries should improve government effectiveness and regulatory quality.

## References

Allen, F., Demirguc-Kunt, A., Klapper, L., & Martinez Peria, M. S. (2016). The foundations of financial inclusion: Understanding ownership and use of formal accounts. *Journal of Financial Intermediation*, *27*(2016), 1–30. https://doi.org/10.1016/j.jfi.2015.12.003

Allen, F., Qian, M., & Xie, J. (2018). Understanding informal financing. *Journal of Financial Intermediation*. https://doi.org/10.1016/j.jfi.2018.06.004

Chong, A., & Calderón, C. (2005). Institutional Quality and Income Distribution. *Economic Development and Cultural Change*, *48*(4), 761–786. https://doi.org/10.1086/452476

Chong, A., & Gradstein, M. (2007). Inequality and informality. *Journal of Public Economics*, *91*(1–2), 159–179. https://doi.org/10.1016/j.jpubeco.2006.08.001

Degryse, H., Lu, L., & Ongena, S. (2016). Informal or formal financing? Evidence on the co-funding of Chinese firms. *Journal of Financial Intermediation*, *27*, 31–50. https://doi.org/10.1016/j.jfi.2016.05.003

Demetriades, P., & Law, S. H. (2006). Finance , Institutions and Economic Development, *260*, 245–260.

Fungáčová, Z., & Weill, L. (2014). Understanding financial inclusion in China. *China Economic Review*, *34*, 196–206. https://doi.org/10.1016/j.chieco.2014.12.004

Islam, A., Nguyen, C., & Smyth, R. (2015). Does microfinance change informal lending in village economies? Evidence from Bangladesh. *Journal of Banking and Finance*, *50*, 141–156. https://doi.org/10.1016/j.jbankfin.2014.10.001

Kaufmann, D., Kraay, A., & Mastruzzi, M. (2010). The Worldwide Governance Indicators: Methodology and Analytical Issues, World Bank Policy Research Working Paper No. 5430. *Available at SSRN: Http://Ssrn. Com/Abstract*, *1682130*(September). https://doi.org/10.1017/S1876404511200046

Khoi, P. D., Gan, C., Nartea, G. V., & Cohen, D. A. (2013). Formal and informal rural credit in the mekong river delta of vietnam: Interaction and accessibility. *Journal of Asian Economics*, *26*, 1–13. https://doi.org/10.1016/j.asieco.2013.02.003

Law, S. H., Azman-Saini, W. N. W., & Ibrahim, M. H. (2013). Institutional quality thresholds and the finance - Growth nexus. *Journal of Banking and Finance*, *37*(12), 5373–5381. https://doi.org/10.1016/j.jbankfin.2013.03.011

Madestam, A. (2014). Informal finance: A theory of moneylenders. *Journal of Development Economics*, *107*, 157–174. https://doi.org/10.1016/j.jdeveco.2013.11.001

Majeed, M. T. (2017). *Quality of Institutions and Inclusive Financial Development in the Muslim World*. *Financial Inclusion and Poverty Alleviation*. https://doi.org/10.1007/978-3-319-69799-4_1

Slesman, L., Zubaidi Baharumshah, A., & Azman-Saini, W. N. W. (2019). Political Institutions and Finance-Growth Nexus in Emerging Markets and Developing Countries: A Tale of One Threshold. *The Quarterly Review of Economics and Finance*. https://doi.org/10.1016/j.qref.2019.01.017

World Bank. (2002). *Building Institutions for Markets*. https://doi.org/ISBN 0-19-521607-5 clothbound ISBN 0-19-521606-7

Zins, A., & Weill, L. (2016). The determinants of financial inclusion in Africa. *Review of Development Finance*, *6*(1), 46–57. https://doi.org/10.1016/j.rdf.2016.05.001